\documentclass[aps,pre,twocolumn,groupedaddress,showpacs,amsmath,amssymb]{revtex4}
\usepackage{epsfig}
\usepackage{amssymb}
\usepackage{amsmath}
\usepackage{graphicx}
\usepackage{amsfonts}
\usepackage{epsfig}
\usepackage{revsymb}
\usepackage{bm}
\usepackage{amsmath}
\usepackage{amssymb}
\usepackage{latexsym}
\usepackage{thumbpdf}
\usepackage{hyperref}

\begin{document}

\title{On the spectrum of facet crystallization waves at the smooth $^4$He crystal
surface}

\author{S. N. Burmistrov}
\email[]{burmi@kurm.polyn.kiae.su} \affiliation{Kurchatov Institute, 123182
Moscow, Russia}


\begin{abstract}
The wavelike processes of crystallization and melting or crystallization waves are
well known to exist at the $^4$He crystal surface in the rough state. Much less is
known about crystallization waves for the $^4$He crystal surface in the smooth
well-faceted state below the roughening transition temperature. To meet the lack,
we analyze here the spectrum of facet crystallization waves and its dependence
upon the wavelength, perturbation amplitude, and the number of possible facet
steps distributed somehow over the wavelength. All the distinctive features of
facet crystallization waves from conventional waves at the rough surface result
from a nonanalytic cusplike behavior in the angle dependence for the surface
tension of smooth crystal facets.
\end{abstract}


\maketitle

\section{Introduction}
\par
Since the prediction by Andreev and Parshin  \cite{An} in 1978, crystallization
waves at the superfluid-solid $^4$He interface have become a well-known
phenomenon. At low temperatures the $^4$He crystal in contact with its liquid
phase can support weakly damped oscillations of the interface due to processes of
crystallization and melting. From the dynamical point of view such weakly damped
crystallization waves at the rough crystal surface are an immediate counterpart of
the familiar gravitational-capillary waves at the interface between two normal
liquids (see review \cite{Ba}).
\par
To date, the crystallization wave phenomena in $^4$He have extensively been
studied for the rough state of crystal surfaces, but not much study has been made
for the well-faceted and atomically smooth crystal surfaces which may have an
infinitely large surface stiffness. The most distinctive feature of the smooth
faceted surface from the rough one is the existence of nonanalytic cusplike
behavior in the angle dependence for the surface tension, e.g., \cite{Ba,La,No}.
The presence of singularity leads to qualitative distinctions in a number of the
phenomena at the smooth faceted crystal surface, e.g., amplitude dependence
velocity of traveling  waves \cite{Gu}, quantum fingering of the inverted
liquid-crystal interface in the field of gravity \cite{Bu}, Rayleigh-Taylor
instability with generation of crystallization waves \cite{Bur}.
\par
In the present work we develop a theory on the spectrum of facet crystallization
waves at the smooth faceted surface of a $^4$He crystal in contact with its liquid
phase. We consider a few possible types of facet crystallization waves and
determine the dispersion relation between the frequency and the wave vector,
perturbation wave amplitude and the number of the crystal facet steps per
wavelength. For simplicity, we discuss the basal plane of hexagonal $^4$He crystal
as an example of the crystal facet and disregard any anisotropy in the basal
plane.

\section{Lagrangian}
\par
Let us assume that the crystal surface is parallel to the $x$-$y$ plane, with
vertical position at $z=0$. In order to derive the oscillation spectrum of a facet
surface, we proceed as follows. Let us call $\zeta =\zeta (t,\, \bm{r})$ the
displacement of the surface from its horizontal position $z=0$. In the lack of
energy dissipation the surface oscillations can be described by the action
\begin{equation}\label{f01}
S=\int dt\, L[\zeta (t,\,\bm{r}),\,\dot{\zeta} (t,\,\bm{r})]
\end{equation}
with the following Lagrangian
\begin{multline*}
 L[\zeta (t,\,\bm{r}),\,\dot{\zeta} (t,\,\bm{r})]=\frac{\rho _\text{ef}}{2}\!\int\! d^2r\,
d^2r'\,\frac{\dot{\zeta}(t,\,\bm{r})\dot{\zeta}(t,\,\bm{r}')}{2\pi
|\bm{r}-\bm{r}'|}
\\
-\!\int\! d^2r\biggl(\alpha (\bm{n})\sqrt{1+(\nabla\zeta )^2}+
\frac{1}{2}\Delta\rho g \zeta ^2 \biggr) .
\end{multline*}
Here $\bm{r}=(x,\, y)$ is a two-dimensional radius-vector. The first term in the
Lagrangian represents the kinetic energy of the interface having an effective
density $\rho _\text{ef}$. We assume that both the liquid and the solid phases are
incompressible. Because of low temperature consideration we will also neglect the
normal component density in the superfluid phase or, equivalently, difference
between the superfluid density $\rho _s$ and the density of the liquid phase
$\rho$. Then the effective interface density $\rho _\text{ef}$ is given by
\begin{equation*}
\rho _\text{ef}=(\rho '-\rho)^2/\rho
\end{equation*}
and depends on the difference  $\Delta\rho =\rho '-\rho$ between the solid density
$\rho '$ and the liquid density $\rho$. For our purposes, the exact magnitude of
the effective density is inessential.
\par
The second and third terms in the Lagrangian are the surface energy and potential
energy of the interface in the field of gravity with acceleration $g$.
\par
Unlike the liquid-liquid interface, the surface tension $\alpha (\bm{n})$ for the
crystal facet depends essentially on the direction of the normal $\bm{n}$ to the
interface. In our simple case this is a function of angle $\vartheta$ alone
between the normal and the crystallographic [0001] or $c$-axis of the crystal hcp
structure with the geometric relation $\mid\tan\theta\mid=\mid\nabla\zeta\mid$.
\par
For the crystal facet tilted by small angle $\vartheta$ from the basal plane, the
expansion of surface tension $\alpha (\theta )$ in series in $\theta$ starts as,
e.g., Refs.\cite{Ba,No},
\begin{equation}\label{f02}
\alpha (\theta ) = \alpha _0+\alpha _1\tan\mid\vartheta\mid +\ldots\, ;
\;\;\;\;\;\; \mid\tan\theta\mid=\mid\nabla\zeta\mid  .
\end{equation}
We do not write the next terms of expansion, e.g., cubic one due to step-step
interaction, since we assume to study only small bending of the crystal surface.
The angular behavior has a nonanalytic cusplike behavior at $\theta =0$ due to
$\alpha _1=\alpha _1(T)$ representing a ratio of the linear facet step energy
$\beta$ to the crystallographic interplane spacing. Below the roughening
transition temperature for the basal plane $T_R\sim$1.2~K the facet step energy
$\beta=\beta (T)$ is positive and vanishes for temperatures  $T>T_R$.
\par
To consider a traveling wave, e.g., propagating to from the left to the right, we
represent the interface perturbation as $\zeta
(t,\,\bm{r})=\zeta(\bm{r}-\bm{V}t)$. Here $\bm{V}$ is the phase velocity of the
wave. Then the action (\ref{f01}) can be written  as
\begin{widetext}
\begin{equation*}
S=\int\! dt\biggl\{\frac{\rho _\text{ef}}{2}\!\int\!\frac{d^2r\, d^2r'}{2\pi
|\bm{r}-\bm{r}'|}\left(\bm{V}\cdot\frac{\partial\zeta
(\bm{r}-\bm{V}t)}{\partial\bm{r}}\right)\left(\bm{V}\cdot\frac{\partial\zeta
(\bm{r}'-\bm{V}t)}{\partial\bm{r}'}\right) -\!\int\! d^2r\biggl(\alpha
(\bm{n})\sqrt{1+(\nabla\zeta )^2}+ \frac{1}{2}\Delta\rho g \zeta ^2
\biggr)\biggr\} .
\end{equation*}
\end{widetext}
Since the integration is performed over $\bm{r}$ and $\bm{r}'$ within the infinite
limits and the kernel in the kinetic term depends on difference
$\mid\bm{r}-\bm{r}'\mid$ alone, we can shift the argument in $\zeta$ by $\bm{V}t$.
Next, by integrating twice by parts, we arrive at
\begin{widetext}
\begin{equation*}
S=\int\! dt\biggl\{\int\! d^2r\, d^2r'\, \frac{\rho _\text{ef}}{2}\,\zeta
(\bm{r})\zeta ( \bm{r}')\left(\bm{V}\cdot
\frac{\partial}{\partial\bm{r}'}\right)\left(\bm{V}\cdot\frac{\partial}{\partial\bm{r}}\right)
\frac{1}{2\pi |\bm{r}-\bm{r}'|} -\!\int\! d^2r\biggl(\alpha
(\bm{n})\sqrt{1+(\nabla\zeta )^2}+ \frac{1}{2}\Delta\rho g \zeta ^2
\biggr)\biggr\} .
\end{equation*}
\end{widetext}
\par
In what follows, we will study sufficiently small bending of the crystal surface
with the sufficiently small displacements $\zeta$ and small angles $\theta$.
Involving inequality $\mid\nabla\zeta\mid\ll1$ and $\mid\tan\theta\mid
=\mid\nabla\zeta\mid$, we take only first terms in the expansion of the surface
energy
 $$
\alpha (\bm{n})\sqrt{1+(\nabla\zeta )^2}\approx \alpha _0+\alpha
_1\mid\nabla\zeta\mid+\alpha _0 (\nabla\zeta )^2/2 .
 $$
Next, we will choose the $x$-axis as a direction of the wave propagation
$\bm{V}=(V,\, 0,\,  0)$ and replace $x'$ with $x$ in the spatial derivatives.
Finally, we arrive at
\begin{widetext}
\begin{equation}\label{f03}
S[\zeta]-S[\zeta =0]=-\int\! dt\bigg\{\,\frac{\rho _\text{ef}V^2}{2}\!\int\!
d^2r\, d^2r'\, \zeta (x)\zeta (x') \frac{\partial^2}{\partial x^2 }\left(
\frac{1}{2\pi |\bm{r}-\bm{r}'|}\right) +\int\! d^2r\biggl(\alpha
_1\mid\nabla\zeta\mid+\frac{\alpha _0}{2}(\nabla\zeta )^2+\Delta\rho g \frac{\zeta
^2}{2}\biggr)\biggr\} .
\end{equation}
\end{widetext}
Variation $\delta S/\delta\zeta (x)$ yields the equation for interface
oscillations
\begin{equation}\label{f04}
\int G(\bm{r}-\bm{r}')\zeta (x')\, d^2r'+\alpha _1\frac{\partial}{\partial
x}\left(\text{sgn}\bigl(\frac{\partial\zeta}{\partial x}\bigr)\right)=0\, .
\end{equation}
Here, for convenience, we have introduced the Green function according to
\begin{multline}\label{f05}
G(\bm{r}-\bm{r}') =-\rho _\text{ef}V^2\frac{\partial^2}{\partial x^2 }\left(
\frac{1}{2\pi |\bm{r}-\bm{r}'|}\right)
\\
+\alpha _0\frac{\partial^2}{\partial x^2 }\delta (\bm{r}-\bm{r}')-\Delta\rho g\, .
\end{multline}
The solution of Eq.~(\ref{f04}) for $V=0$ has been studied in Ref.\cite{Bu}.
\par
Before solving Eq.~(\ref{f04}), we make the following remarks. First, in the
regions with $\zeta '(x)\neq 0$ the equation~(\ref{f04}) reduces to a linear
equation with the difference kernel
 $$
\int G(\bm{r}-\bm{r}')\zeta (x')\, d^2r' =0\, .
 $$
The solution can be found as a sum of independent Fourier harmonics $\zeta
(x)=\sum _q\zeta _q\exp (iqx)$. For a single harmonic
\begin{equation}\label{f06}
\zeta (x)=\zeta _q\exp (\pm iqx)\;\;\text{or}\;\;\zeta (x)=\zeta _q\cos q(x-x_0) ,
\end{equation}
one should have $G(q)\zeta_q=0$. For an existence of nontrivial solutions, it is
necessary to put $G(q)=0$. Thus, vector $q$ must satisfy the condition
\begin{equation}\label{f07}
G(q)=\rho _\text{ef}\, V^2\frac{q^2}{q}-\alpha _0q^2-\Delta\rho g=\rho
_\text{ef}\, q\bigl( V^2 -V_0^2(q)\bigl)=0.
\end{equation}
Here we have introduced notation $V_0(q)$ for the phase velocity of
crystallization waves at the rough crystal surface with the spectrum $\omega
_0(q)$:
\begin{equation*}
V_0(q)=\omega _0(q)/q\, , \;\;\;\; \omega _0(q)= \sqrt{(\alpha _0q^3+\Delta\rho
gq)/\rho _\text{ef}\,}\, .
\end{equation*}
\par
Second, in the regions with $\zeta '(x)\equiv 0$ the solution is trivial, i.e.,
 $$
\zeta (x)=\text{const} .
 $$
And the last, since during the melting-crystallization process the total mass of
the solid and liquid phases remains invariable, the solution of Eq.~(\ref{f04})
must satisfy the following condition
 $$
\int \zeta (x)\, dx =0 .
 $$
 \par
\begin{figure}
\includegraphics[scale=0.6]{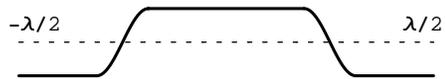}
\caption{The wave is formed by alternating one flat segment of length $(\lambda
-l)/2$ and kink of width $l/2$.} \label{fig1}
\end{figure}
\begin{figure}
\includegraphics[scale=0.6]{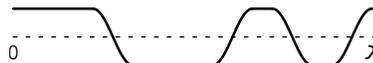} \caption{The wave is composed of alternating two flat
segments of different lengths and four kinks of width $l/2$ each.} \label{fig2}
\end{figure}

As a result, the general solution for profile $\zeta (x)$ should represent a train
of flat segments and half-sinusoids. The width of  a half-sinusoid, which we
denote $l/2$, is governed by the magnitude of vector $q$ according to $G(q)=0$ or
$V_0(q)=V$ and is equal to  $l/2=\pi /q$. The half-sinusoid, which connects two
neighboring flat segments, can be regarded as a macroscopic facet step in contrast
to elementary steps of an atomic scale. To illustrate, we give two examples of
such facet crystallization waves with alternation of one or two various flat
segments in Figs.~\ref{fig1} and~\ref{fig2}.
It is obvious that the wavelength $\lambda$ cannot be less than a sum of two
half-sinusoids $l=2\pi /q$. In other words, the wave number $k=2\pi /\lambda$ does
not exceed vector $q$, i.e., $k\leqslant q$. Later, it will be seen that always
$\omega (k)\geqslant\omega _0(k)$.
\par
Besides the wavelength, the wave spectrum at the crystal facet will also be
characterized by the number of flat segments and their amplitudes. The width of
macroscopic crystal steps $l/2=\pi /q$ depends on a ratio of wave frequency
$\omega$ to wave number $k$ and can be determined using relation
\begin{equation}\label{f07a}
\frac{\omega}{k}=\frac{\omega _0(q)}{q}\, .
\end{equation}
Provided $l\ll\lambda$, the width of half-sinusoids can be neglected in first
approximations and the macroscopic facet steps can be treated as a kink of zero
width. Usually, this is a range of sufficiently high frequencies $\omega
(k)\gg\omega _0(k)$ and small wave perturbation amplitudes
$\mid\!\zeta\!\mid\,\alt l \ll\lambda$.

\section{Simplest Type of Waves}
\par
Let us start from the simplest type of the crystallization wave which is formed by
alternating a single flat segment of height $\zeta$ and a half-sinusoid of width
$l/2$ so that $\int\zeta(x)\, dx=0$ (Fig.~\ref{fig1}). In this case the wave
perturbation is specified by
\begin{eqnarray}
\zeta (x)=\zeta\left\{
\begin{array}{cc}
1, & \mid x \mid <\frac{\lambda -l}{4}
\\
\\
\!\cos\left[\frac{2\pi}{l}\bigl(\mid x\mid -\,\frac{\lambda -l}{4}\bigr)\right], &
\frac{\lambda -l}{4} <\mid x\mid <\frac{\lambda +l}{4}\, .
\\
\\
-1, & \frac{\lambda +l}{4} <\mid x\mid < \frac{\lambda}{2}
\end{array}
\right.
\end{eqnarray}
The total length of the flat segments is equal to $\lambda -l$. For correctness of
approximation used, we should assume that $\mid\zeta\mid\ll
l/2\pi\leqslant\lambda/2\pi$.
\par
To determine the spectrum for such type of the facet wave, we calculate the action
as a function of the wave amplitude $\zeta$ and then minimize the action. The
calculation is readily performed in the Fourier representation using
\begin{multline}\label{f09}
\zeta (x)=\sum _n\zeta _ne^{ik_nx} ,
\\
k_n=2\pi n/\lambda =kn \;\;\;\; (n=0,\,\pm 1,\,\pm 2\ldots) .
\end{multline}
Hence we have for the variation of the action taken from the flat crystal surface
and ascribed per wavelength
\begin{equation}\label{f10}
S[\zeta]-S[\zeta =0]= \frac{1}{\lambda}\sum\limits
_{n=-\infty}^{\infty}G_{\omega}(k_n)\zeta _n\zeta _{-n}-4\alpha _1\!\mid\zeta\mid
.
\end{equation}
The second term is a contribution due to formation of the facet segment with the
step amplitude $2\zeta$. The number of such steps over wavelength  equals 2 in our
specific case. The calculation of Fourier components is straightforward
\begin{multline}\label{f11}
\zeta _n=\int _{\lambda /2}^{\lambda /2}\zeta (x)e^{-ik_nx}\, dx = \zeta\lambda
f_n\, ,
\\
f_n =\frac{\sin(\pi n/2)}{(\pi n/2)}\left(\frac{\cos (\pi nl/2\lambda
)}{1-n^2l^2/\lambda ^2}-\cos\frac{\pi n}{2}\right)
\end{multline}
with the obvious properties $\zeta _{-n}=\zeta _n$ and $\zeta _{n=2m}=0$. For
$\lambda =l$, harmonics $f_1=f_{-1}=1/2$ alone remain nonzero. Inserting
Eq.~(\ref{f11}) into (\ref{f10}) and minimizing the action $\partial
S/\partial\!\mid\!\zeta\mid =0$, we arrive at the equation which determines the
spectrum of facet crystallization wave
\begin{equation}\label{f12}
\sum _n G_{\omega}(k_n)f_n^2=\frac{2\alpha _1}{\lambda \mid\zeta\mid }\, .
\end{equation}
Then, using (\ref{f11}) and $l=2\pi /q$, we have
\begin{widetext}
\begin{multline}\label{f13}
\sum\limits _{m=0}^{\infty}\bigl(\frac{\rho _\text{ef}\omega
^2}{k}\,\frac{1}{2m+1} -\alpha _0k^2(2m+1)^2-\Delta\rho
g\bigr)f_{2m+1}^2=\frac{\alpha _1k}{2\pi\mid\zeta\mid} \, ,
\\
f^2_{2m+1}=\frac{4}{\pi ^2}\,\frac{1}{(2m+1)^2}\,\frac{\cos
^2\bigl[\frac{\pi}{2}(2m+1)\frac{k}{q}\bigl]}{\bigl[1-(2m+1)^2k^2/q^2\bigr]^2}\, ,
\;\;\;\; m=0,\, 1,\, 2,\ldots
\end{multline}
\end{widetext}
The magnitude of vector $q$ is determined from the condition $\omega
_0(q)/q=\omega /k$.
\par
We first analyze the limiting case of infinitely narrow step $l\ll\lambda$ or
$k\ll q$.  In this approximation one can here neglect the contributions from the
regular $\alpha _0$ surface and gravitational $\Delta\rho g$ terms and estimate
$f_{2m+1}$ at point $k=0$. Then, a sum in (\ref{f13}) reduces to
 $$
\rho _\text{ef}\,\frac{\omega ^2}{k}\,\frac{4}{\pi ^2}\sum\limits
_{m=0}^{\infty}\,\frac{1}{(2m+1)^3}=\frac{7\zeta (3)}{2\pi ^2}\, \frac{\omega
^2}{k}\, ,
 $$
where $\zeta (3)\approx 1.20$ is the Riemann zeta-function. Finally, we get the
spectrum
\begin{equation}\label{f14}
\omega ^2=\frac{\pi}{7\zeta (3)}\,\frac{\alpha _1k^2}{\rho
_\text{ef}\mid\zeta\mid}=\frac{4\pi ^3}{7\zeta (3)}\,\frac{\alpha _1}{\rho
_\text{ef}\,\lambda ^2\mid\zeta\mid}\, .
\end{equation}
The phase velocity $V$ depends on the perturbation amplitude alone
 $$
V=\frac{\omega}{k}=\left(\frac{\pi}{7\zeta (3)}\,\frac{\alpha _1}{\rho
_\text{ef}\mid\zeta\mid}\right)^{1/2} .
 $$
A special feature of the spectrum is its growing stiffness as
$\mid\zeta\mid\rightarrow 0$ when the frequency of surface oscillations becomes
infinitely large. Regardless of magnitudes $\alpha _0$ and $\Delta\rho g$ the
behavior $\omega\sim\mid\zeta\mid ^{-1/2}$ is universal in the
$\mid\zeta\mid\rightarrow 0$ limit.
\par
To understand the typical magnitudes and applicability of our approximations, we
start from the case when wavelength $\lambda$ exceeds slightly the double width of
crystal facet step, i.e., $(\lambda -l)/\lambda\ll 1$. Then, in Eq.~(\ref{f13})
the main contribution results from the first term with $m=0$. Approximately, one
has
\begin{equation}\label{f15}
\omega (k)= \omega _0(k)+\frac{1}{\pi}\,\frac{\alpha _1k^2}{\rho
_{\text{ef}}\,\omega _0(k)}\,\frac{1}{\mid\zeta\mid}\, .
\end{equation}
The effect of crystal facet smoothness on the wave spectrum due to nonzero $\alpha
_1$ is governed by a ratio
 $$
\frac{\alpha _1 k^2}{\alpha _0 k^2+\Delta\rho
g}\,\frac{1}{k\mid\zeta\mid}=\frac{2\pi\alpha _1/\alpha _0}{1+\lambda ^2/\lambda
_0^2}\,\frac{\lambda}{\mid\zeta\mid}\, ,
 $$
where $\lambda _0=2\pi\sqrt{\alpha _0/\Delta\rho g}$ is a usual capillary length.
To have a strong effect on the spectrum in the sense $\omega (k)\gg\omega _0(k)$,
we should satisfy the inequality
\begin{equation}\label{f16}
\mid\zeta\mid\ll \frac{2\pi\alpha _1}{\alpha _0}\,\frac{\lambda}{1+\lambda
^2/\lambda _0^2}\, .
\end{equation}
\par
Below, in detail, we will analyze most interesting region of sufficiently small
perturbation amplitudes $\zeta\rightarrow 0$ and wavelengths smaller than
capillary length $\lambda\ll\lambda _0$. Neglecting gravitational term in
(\ref{f13}), we have
\begin{multline*}
\sum\limits _{m=0}^{\infty}\left(\frac{1}{(2m+1)^3} - \,\frac{k^3}{q^3}
\right)\frac{\cos
^2\bigl[\frac{\pi}{2}(2m+1)\frac{k}{q}\bigl]}{\bigl[1-(2m+1)^2k^2/q^2\bigr]^2}
\\
=\frac{\pi}{8}\,\frac{\alpha _1k^2}{\rho_{\text{ef}}\,\omega ^2\!\mid\zeta\mid} \,
.
\end{multline*}
Involving that $k/q =\alpha _0k^3/(\rho_{\text{ef}}\,\omega ^2)\ll 1$ and
estimating the above sum within logarithmic accuracy as
 $$
7\zeta (3)/8- (\pi ^2\!/8\, -1)(k/q)^2\ln (q/k) ,
 $$
we find the spectrum with the correction due to finiteness of $\alpha _0$
\begin{multline*}
\omega ^2\approx \frac{\pi}{7\zeta (3)}\,\frac{\alpha
_1k^2}{\rho_{\text{ef}}\!\mid\zeta\mid}\left[1+\frac{\pi ^2-8}{21\zeta
(3)}\biggl(\frac{7\zeta (3)}{\pi}\,\frac{\alpha _1}{\alpha _0}\,
k\!\mid\zeta\mid\biggr)^{2/3}\right.
\\
\left. \times\ln\biggl(\frac{\pi\alpha _1}{7\zeta(3)\alpha
_0}\,\frac{1}{k\!\mid\zeta\mid}\biggr)\right] .
\end{multline*}
The width of the kink between two flat segments can readily be estimated from
$q=\rho _{\text{ef}}\,\omega ^2(k)/(\alpha _0k^2)$ as
\begin{equation}\label{f17}
l/2\approx 7\zeta (3)\frac{\alpha _0}{\alpha _1}\,\mid\zeta\mid .
\end{equation}
Thus, the approximation of zero-width facet step $l\ll\lambda$ can be justified
for the small amplitude perturbations if
 $$
\mid\zeta\mid\ll\frac{\lambda}{14\zeta (3)}\,\frac{\alpha _1}{\alpha _0}\, .
 $$
\par
On the whole, the spectrum of facet crystallization waves can qualitatively be
described by introducing effective surface tension or stiffness dependent on both
wave vector and perturbation amplitude \cite{Bu}
\begin{equation}\label{f18}
\alpha _{\text{ef}}\rightarrow \alpha _0+\frac{\pi}{7}\,\zeta (3)\frac{\alpha
_1}{k\!\mid\zeta\mid}\, .
\end{equation}

\section{Waves With a Few Crystal Facet Steps}
\par
\begin{figure}
\includegraphics[scale=0.7]{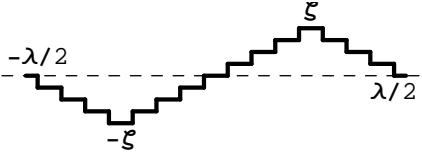}
\caption{The sketch of crystallization wave of the total perturbation amplitude
$\zeta$ and wavelength $\lambda$.}\label{fig3}
\end{figure}

Here we consider another type of facet crystallization waves with an arbitrary
number of crystal facet steps per wavelength $\lambda$. The wave amplitude $\zeta$
is assumed to be sufficiently small in order to neglect the regular surface
$\alpha _0$ and gravitational $\Delta\rho g$ terms. This limit corresponds to high
frequencies $\omega (k)\gg\omega _0(k)$ and zero width of the kink between two
flat segments. Thus, the profile of the perturbed crystal surface represents a
broken line consisting of vertical steps and horizontal segments. For the
definiteness, we consider the surface profile from $4N$ vertical steps linking the
same $M=4N$ flat facet segments with length $\Delta x=\lambda /M$
(Fig.~\ref{fig3}).
The vertical steps, having the same height of $\zeta /N$, are located at the
points
  $$
x_k=\pm\frac{2k-1}{2}\Delta x , \;\;\; k=1,\, 2, \ldots\, , \, N .
 $$
The points $x_{k+1}$ and $x_k$ are connected with a horizontal segment of length
$\Delta x$ with the vertical amplitude
 $$
\zeta _k=\pm\zeta\,\frac{N-\mid k-N\mid}{N}\, ,\;\;\; k=0,\, 1,\, 2,\ldots , \, 2N
.
 $$
In accordance with (\ref{f09}) we find the Fourier components for $\zeta (x)$
\begin{multline*}
\zeta _n =\int\limits _{-\lambda /2}^{\lambda /2}\!\! dx\,\zeta (x)e^{-2\pi
inx/\lambda}
\\
=-2i\sum\limits _{k=1}^{2N-1}\zeta ^{\frac{N-\mid k-N\mid}{N}}\int\limits
_{x_k}^{x_{k+1}}\!\sin\bigl(2\pi nx/\lambda\bigr)\, dx
\\
=-i\frac{i\lambda}{\pi n}\,\frac{\zeta}{N}\sum\limits _{k=1}^{N}\bigl(\cos
y_k\,-\cos y_{k+N}\bigr),\;\;\; y_k=\frac{\pi n}{4N}(2k-1).
\end{multline*}
Summation, in essence, is reduced to summing geometric series. As a result, we
obtain
\begin{equation}\label{f19}
\zeta _n=\zeta\lambda f_n\, , \;\;\;\; f_n=\frac{i}{2\pi n}\,\frac{1-(-1)^n}{N}\,
\frac{\sin (\pi n/2)}{\sin (\pi n/4N)}
\end{equation}
with the obvious properties $\zeta _{-n}=\zeta _n$ and $\zeta _{n=2m}=0$.
\par
Unlike derivation of Eq.~(\ref{f10}), we must take here into account that each
crystal step contributes $\alpha _1\mid\zeta\mid\! /N$ into the action and the
number of steps equals $4N$ at the wavelength. Since $4N\alpha _1\mid\zeta\mid\!\!
/N=4\alpha _1\!\mid\zeta\mid$, we have the same contribution to the action from
the steps and thus the same form (\ref{f12}) of the equation to determine the wave
spectrum
\begin{equation}\label{f20}
\sum\limits _{m=0}^{\infty}\frac{\rho _\text{ef}\,\omega ^2}{k} \,\frac{\mid
f_{2m+1}\mid ^2}{(2m+1)}=\frac{\alpha _1k}{2\pi\mid\zeta\mid}
\end{equation}
but with another Fourier component $f_{2m+1}$. Applying $f_{2m+1}$ from
(\ref{f19}), we have
\begin{equation*}
\omega^2\sum\limits _{m=0}^{\infty}\frac{1}{(2m+1)^3}\,\frac{1}{N^2\sin ^2[\pi
(2m+1)/4N]}= \frac{\pi}{2}\,\frac{\alpha _1k^2}{\rho _{\text{ef}}\mid\zeta\mid}\,
.
\end{equation*}
\par
Finally, the spectrum of facet crystallization waves is determined by
\begin{equation}\label{f21}
\omega _N=s_N\omega _1\, , \;\;\;\; \omega _1^2=\frac{2\pi}{7\zeta
(3)}\,\frac{\alpha _1k^2}{\rho _{\text{ef}}\mid\zeta\mid}\, ,
\end{equation}
where $s_N$ is given by
 $$
s_N^2=\frac{4\sum\limits _{m=0}^{\infty}(2m+1)^{-3}}{N^2\sum\limits
_{m=0}^{\infty}(2m+1)^{-3}\sin^{-2}[\pi (2m+1)/4N]}
 $$
and
 $$
s_{\infty}^2=\frac{7\pi ^2\zeta (3)}{62\zeta (5)}\, .
 $$
The factor $s_N$ varies insignificantly within the range from 1 for $N=1$ to 1.136
for $N=\infty$ (Table~\ref{tab1}).
\begin{table}
\caption{The ratio of frequencies $s_N=\omega _N/\omega _1$ for facet
crystallization waves with the different number $N$ of crystal facet steps per
wavelength}
\label{tab1}
\begin{ruledtabular}
\begin{tabular}{ccccccccc}
$N$ & 1 & 2 & 3 & 4 & 5 & 6 & 7 & $\infty$
\\
$s_N$ & 1.000 & 1.103 & 1.122 & 1.128 & 1.131 & 1.133 & 1.134 & 1.136
\end{tabular}
\end{ruledtabular}
\end{table}
\par
We have analyzed above the wave spectrum for the regular arrangement and identical
height of crystal steps over wavelength. This is not, of course, solely possible
structure with $4N$ crystal steps. The location of crystal steps and their heights
can have an arbitrary and disordered structure. However, the dimensional estimate
(\ref{f21}) for the spectrum holds for. As concerns the factor $s_N$, it varies
slightly as a function of the perturbation profile.
\par
Provided the regular part of surface tension $\alpha _0$ differs from zero, the
maximum number of crystal steps $4N_m$ is limited. Using magnitude $l$ for the
width of the  kink between two flat segments
  $$
l\sim \frac{\alpha _0}{\alpha _1}\,\mid\zeta\mid ,
 $$ we estimate the
number $4N_m$ of possible crystal steps for the given frequency $\omega$ according
to
\begin{equation}\label{f22}
4N_m\sim\frac{\lambda}{l}\sim\frac{\alpha _1}{\alpha
_0}\,\frac{\lambda}{\mid\zeta\mid}\, .
\end{equation}
Obviously, the smaller the perturbation amplitude $\mid\zeta\mid$, the larger the
number of possible crystal steps.
\par
On the whole, the frequency of crystallization waves at a smooth crystal facet
proves to be dependent not only on the wavelength, but also on the wave amplitude
and the number of crystal facet steps which can be placed within the wavelength.
From the experimental point of view this means that the excitation of
crystallization waves with a fixed frequency will result in exciting some train of
waves with different wavelengths, amplitude, and the number of crystal facet
steps. In this connection the shape of the perturbed crystal facet will resemble
rather an irregular and ill-defined profile with some elements of irregular-like
character.  A weak dependence of the wave frequency on the number of steps
facilitates such phenomenon.  In some sense one might say about transition to a
rough state of the surface and destruction of the crystal faceting \cite{Bur}.

\section{Conclusion}
\par
The crystallization waves at the smooth crystal facets are expected to demonstrate
a more varied and complicated picture than that at the rough crystal surfaces. The
plane crystallization wave represents an alternation of flat crystal facets linked
via macroscopic crystal steps with the width dependent on the wave velocity. Most
striking phenomena should appear in the limit of sufficiently small perturbation
amplitudes. The frequency spectrum of facet waves depends significantly on the
perturbation amplitude. The dependence on the structure of wave perturbation and
the number of crystal facet steps is not so drastic. Excitation of waves at a
given frequency should produce a train consisting of waves differing in
wavelength, structure and the number of facet steps and interacting  nonlinearly
with each other. This all is in contrast to harmonic waves which exist at the
rough crystal surfaces and have the lower frequencies at the same wavelengths.
Evidently, the distinction results from a singularity in the angular behavior of
the surface tension for the smooth crystal facets.
\par
Let us estimate typical frequencies for the short wavelength range
$\lambda\ll\lambda _0\sim$ 6~mm. Taking $\rho _{\text{ef}}\sim$ 2~mg/cm$^3$,
$\alpha _1\sim$ 0.014~erg/cm$^2$ and $\alpha _0\sim$ 0.16--0.18~erg/cm$^2$, e.g.,
\cite{Ba} for the (0001) $^4$He facet, we have the frequency $\omega\sim$ 10~kHz
and velocity $V\sim$ 2~m/s for the wavelength $\lambda\sim$ 1 mm and perturbation
amplitude $\mid\zeta\mid\sim$ 1~$\mu$m. In this case one may expect the maximum
number of possible steps over wavelength to $\sim$ 100. If the perturbation
amplitude for the same $\lambda$ approaches $\mid\zeta\mid\sim$ 0.1~mm, only one
or two steps become possible. The wave frequency reduces to about $\omega\sim$
2~kHz which insignificantly exceeds the magnitude at the rough surface. Note that,
for the perturbation amplitude of a crystal lattice spacing in height, the
propagation velocity $V$ reaches the magnitudes of $\sim$100~m/s comparable with
the sound velocity.
\par
In principle, one can find a few casual mentions about phenomena similar to the
formation of crystallization waves at the crystal $^4$He facets under heavy shake
of an experimental cell \cite{Ke} or in the process of anomalously fast growth of
a $^4$He crystal under high overpressures \cite{Ts,Tsy}. More convincing
observation in favor of an existence of progressive facet waves has recently been
found \cite{Tse} at the crystal (001) facet in $^3$He. Apparently, one of
complicating factors in exciting and studying facet crystallization waves may be
associated with the threshold character for most of phenomena occurring at the
smooth crystal facets. In particular, it may require a sufficiently large size of
the facet and sufficiently high amplitudes of driving perturbation. In this
connection it may be helpful to employ the conditions close to an onset of some
instability, e.g., electrocapillary one in an electric field across the
interface\cite{Bu}, Rayleigh-Taylor \cite{Bur} or Faraday instabilities \cite{Ab}.
\newline
\par
The work is supported in part by the RFBR Grants Nos. 08-02-000752a and
10-02-00047a.


\begin{thebibliography}{99}
\bibitem{An} A.~F.~Andreev and A.~Ya.~Parshin,  Zh. Eksp. Teor. Fiz. \textbf{75}, 1511 (1978)
[Sov. Phys. JETP \textbf{48}, 763 (1978)].
\bibitem{Ba}S. Balibar, H. Alles, and A. Ya. Parshin, Rev. Mod. Phys. \textbf{77}, 317 (2005).
\bibitem{La} L. D. Landau, \textit{The Equilibrium Form of Crystals}, in Collected Papers (Pergamon, Oxford,
1965).
\bibitem{No}P. Nozi\`{e}res, in \textit{Solids Far From Equilibrium}, edited by C.Godr\`{e}che (Cambridge
University Press, Cambridge, 1992), p. 38.
\bibitem{Gu} R. B. Gusev and A. Ya. Parshin, Pisma ZhETF \textbf{85}, 717 (2007) [JETP Lett. \textbf{85}, 588
(2007)].
\bibitem{Bu}S. N. Burmistrov and L. B. Dubovskii, {\it J. Low Temp. Phys.} \textbf{150}, 295 (2008).
\bibitem{Bur}S. N. Burmistrov, L. B. Dubovskii, and  V. L. Tsymbalenko, Phys. Rev. E \textbf{79}, 051606 (2009).
\bibitem{Ke} K. O. Keshishev, A. Ya. Parshin, and A. I. Shalnikov, in \textit{Soviet Scientific Reviews,
Sec. A: Physics Reviews}, edited by I. M. Khalatnikov (Harwood Academic, New York,
1982), vol. \textbf{4}, p. 155.
\bibitem{Ts} V.~L.~Tsymbalenko, Phys. Lett. A \textbf{274}, 223 (2000).
\bibitem{Tsy} V.~L.~Tsymbalenko,  Zh. Eksp. Teor. Fiz. \textbf{126}, 1391 (2004) [JETP \textbf{99}, 1214 (2004)].
\bibitem{Tse}V. Tsepelin, H. Alles, A. Babkin, R. Jochemsen, A.~Ya.~Parshin, and
I.~A.~Todoshchenko, {\it J. Low Temp. Phys.} \textbf{129}, 489 (2002), p. 525.
\bibitem{Ab} H.~Abe, T.~Ueda. M.~Morikawa, Yu~Saitoh, R.~Nomura, and Y.~Okuda, J. Phys. Conf. Ser.
\textbf{92}, 012157 (2007).
\end{thebibliography}
\end{document}